# Human Factors in Detecting AI-Generated Portraits: Age, Sex, Device, and Confidence

Sunwhi Kim[1 †], Sunyul Kim[2]

Hwasung Medi-Science University, Dept. of Bio-Healthcare[1]

Yonsei University, Graduate School of Engineering, Dept. of Artificial Intelligence[2]

## Abstract

Generative AI now produces photorealistic portraits that circulate widely in social and newslike contexts. Human ability to distinguish real from synthetic faces is time-sensitive because image generators continue to improve while public familiarity with synthetic media also changes. Here, we provide a time-stamped snapshot of human ability to distinguish real from AI-generated portraits produced by models available in July 2025. In a large-scale web experiment conducted from August 2025 to January 2026, 1,664 participants aged 20-69 years (mobile n = 1,330; PC n = 334) classified one portrait per trial as REAL or AI. Each participant judged 20 trials sampled from a 210-image pool comprising real FFHQ photographs and AI-generated portraits from ChatGPT-4o and Imagen 3. Overall accuracy was high (mean 85.2%, median 90%) but varied across groups. PC participants outperformed mobile participants by 3.65 percentage points. Accuracy declined with age in both device cohorts and more steeply on mobile than on PC (-0.607 vs -0.230 percentage points per year). Self-rated AI-detection confidence and AI exposure were positively associated with accuracy and statistically accounted for part of the age-related decline, with confidence accounting for the larger share. In the mobile cohort, an age-related sex divergence emerged among participants in their 50s and 60s, with female participants performing worse. Trial-level reaction-time models showed that correct AI judgments were faster than correct real judgments, whereas incorrect AI judgments were slower than incorrect real judgments. ChatGPT-4o portraits were harder and slower to classify than Imagen 3 portraits and were associated with a steeper age-related decline in performance. These findings frame AI portrait detection as a human-factors problem shaped by age, sex, device context, and confidence, not image realism alone.

## Introduction

Generative AI now produces portrait images with sufficient photorealism to circulate as profile pictures, advertisements, social media posts, and even news-like visuals. This raises a practical and scientific question: can people reliably distinguish real photographs from AI-generated portraits? The answer matters for fraud detection, credibility judgments, and the design of user-facing safeguards.

Research on deepfake and synthetic-media detection has generally found that human performance is limited. Across modalities, detection is often modest and sometimes near chance, especially for high-quality content, and several studies suggest a bias toward accepting material as authentic[1–4] . Köbis et al. (2021)[1] showed that people could not reliably detect deepfakes and did not improve simply through awareness or financial incentives. Groh et al. (2022)[2] likewise found asymmetric errors, with participants identifying real videos more accurately than deepfakes and tending to favor "real" responses. Extending beyond video, Frank et al. (2024)[4] reported that state-of-the-art generated image, audio, and text samples were often nearly indistinguishable from human-generated media across countries.

Synthetic faces are a particularly important case because faces are central to identity, trust, and social inference. In GAN-era face synthesis, observers often failed to distinguish synthetic from real faces, and AI-synthesized faces were even judged as more trustworthy than real ones[5]. Subsequent work introduced the concept of AI hyperrealism, showing that some AI faces—especially White AI faces—can be judged as human more often than actual human faces[6]. These findings suggest that "uncanny valley" intuitions are not a dependable defense against contemporary face synthesis.

At the same time, detection is not a fixed property of either generators or observers. Generator technology evolves rapidly, and cues that were diagnostic for earlier GAN-based systems may not transfer to newer diffusion and multimodal image models. Evidence on improvement is also mixed: passive cue lists may offer little benefit, whereas stronger interventions such as feedback-based training or AI assistance can improve performance under some conditions[3,7]. A useful reframing, therefore, is that AI detection is a moving target. As generators become more realistic, people may also accumulate exposure, learn new cues, and recalibrate their expectations about what AI-generated content looks like.

From this perspective, the key question is not only whether people can discriminate AI portraits from real photographs, but how that ability is distributed across the population and which human factors shape it. The present study addresses this question with a time-stamped snapshot of human performance in detecting AI-generated portraits produced by models available as of July 2025, using behavioral data collected from August 2025 through January 2026. Our aim is not to establish a permanent benchmark. Rather, we seek to characterize what was detectable at that moment, how detection performance varied across individuals and contexts, and which human factors predicted better discrimination.

To do so, we conducted a large-scale web-based experiment in which participants judged one portrait at a time as REAL or AI (a single-interval yes/no task), using real face portraits and AI-generated counterparts produced by two leading systems available at the time of stimulus creation: ChatGPT-4o (native image generation) and Google Imagen 3 (via Gemini 2.5). AI portraits were created using a mirroring procedure designed to promote high-fidelity synthesis (see Methods). In addition to accuracy and reaction time, we measured self-reported AI exposure frequency, AI-detection confidence, and decision strategies.

This design allows us to move beyond an overall accuracy estimate in several ways. First, we test whether relatively high mean accuracy can coexist with age-related decline and whether such decline depends on device context. Second, we examine whether AI exposure and AI-detection confidence predict accuracy in ways consistent with adaptation, and whether they partly mediate age-related differences. Third, we explore whether these age-related patterns differ by sex. Finally, we analyze reported decision cues and trial-level response times to ask whether correct and incorrect judgments follow different processing patterns for real versus AI-generated portraits.

Together, this study treats detection of AI-generated portraits as a historically situated human-factors problem shaped by model evolution, real-world exposure, and individual adaptation. By anchoring stimuli to July 2025 and behavior to late 2025–early 2026, we offer a dated map of the human–AI detection landscape and of the demographic, contextual, and experiential gradients that structure performance within it.

## Results

### Overview of sample and task performance

**Figure 1** summarizes the stimulus set and overall task flow. Participants completed a web-based REAL/AI classification task comprising 20 main trials, each presenting a single portrait, sampled from a pool of 210 portraits (70 real FFHQ, 70 ChatGPT-4o, and 70 Imagen 3), followed by brief self-report measures. A total of 1,843 records were collected in the public web experiment. After applying prespecified filters (first attempt only; age 20–69), the final analytic sample included 1,664 participants (**Fig. 2A**). Overall accuracy across included participants showed a broad distribution, with a median of 90 and a mean of 85.24 (**Fig. 2B**). Device composition was strongly skewed toward mobile participation (mobile: n = 1,330; PC: n = 334; **Fig. 2C**; **Supplementary Fig. S0**). Participants on PC achieved higher accuracy than

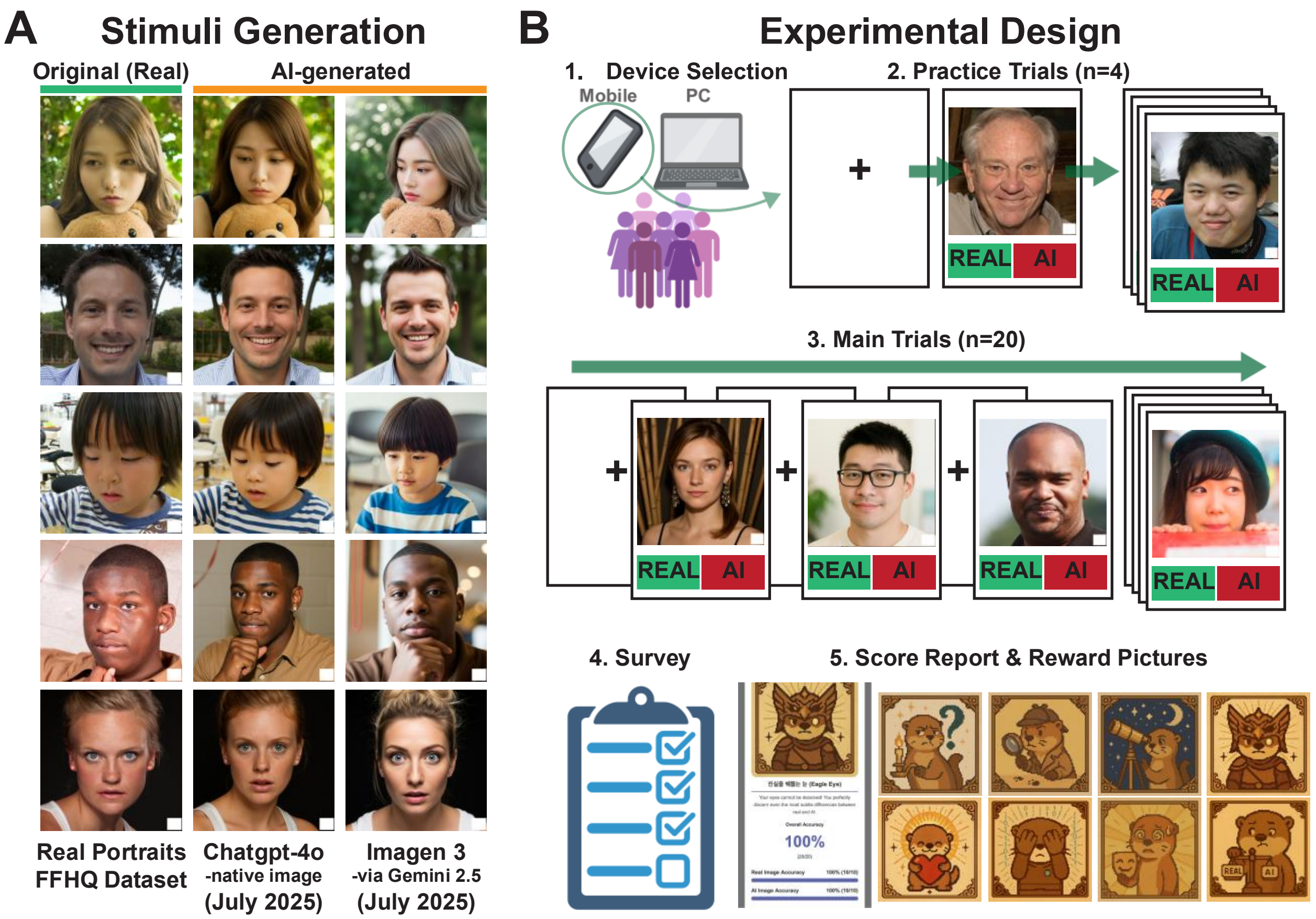

**Figure 1. Stimulus generation and experimental procedure**

(A) Example real portraits and AI-generated portraits used in the experiment. Real portraits were sampled from the Flickr-Faces-HQ (FFHQ) dataset, while AI portraits were generated using state-of-the-art models available at the time of stimulus creation (ChatGPT-4o, native image generation; Google Imagen 3 via Gemini 2.5). Stimuli were created in July 2025.

(B) Overview of the online experimental flow: participants selected their device (mobile or PC), completed practice trials (n=4), then performed main trials (n=20) in which they classified each portrait as real or AI-generated. Feedback was provided only during practice trials; answers were not revealed during the main trials. After the task, participants completed a brief survey assessing AI-related self-report measures (e.g., exposure, confidence, attitude) and reported the strategies used for AI detection. A score report and reward images were presented at the end to encourage engagement.

Real portraits in (A-B) were sampled from FFHQ (Karras et al., 2019). Image-level attribution, license information, and modification notes for the FFHQ stimuli are provided in Supplementary Table S1.

those on mobile (Welch's t-test: $t(697.19) = -5.56$, $p < .001$; $\Delta = 3.65$ percentage points [PC − mobile]), motivating device-stratified analyses throughout.

### Device-dependent age effects on discrimination accuracy

Accuracy declined with age in both device cohorts (**Fig. 2E–F**). In the mobile cohort, age was strongly negatively associated with accuracy (Pearson $r = -0.494$, $p < .001$), and a linear regression estimated a decline of −0.607 pp/year. In the PC cohort, the association was weaker but still significant (Pearson $r = -0.211$, $p < .001$; regression slope = −0.230 pp/year). Slope comparisons indicated a significantly steeper age-related decline on mobile than on PC (Δslope [Mobile − PC] = −0.378 pp/year; $Z = 5.78$, $p < .001$; **Fig. 2G**).

### AI-related self-reports correlate with accuracy and statistically mediate age effects in the mobile cohort

In the mobile cohort, accuracy showed systematic relationships with AI-related self-reports (**Fig. 3A–C**). Specifically, accuracy was positively correlated with AI-detection confidence (Spearman $\rho = 0.408$, $p < .001$) and AI exposure ($\rho = 0.253$, $p < .001$), whereas attitude toward AI showed no reliable association with accuracy ($\rho = -0.010$, $p \geq .05$; **Fig. 3A**).

To test whether these factors statistically accounted for age-related declines in accuracy, we fit a parallel mediation model with age as predictor and exposure, confidence, and attitude as parallel mediators (standardized β; **Fig. 3D**). Indirect effects were considered significant when the bootstrap 95% confidence interval (CI) excluded zero. The total indirect effect was significant (ind_total = −0.111, 95% bootstrap CI [−0.137, −0.087]), with the confidence-mediated pathway dominating (ind2 = −0.085, CI [−0.106, −0.065]) and a smaller but significant exposure pathway (ind1 = −0.026, CI [−0.043, −0.010]); the attitude pathway was not significant (ind3 ≈ 0; CI includes 0). A substantial direct age effect remained after accounting for mediators (c' = −0.397, CI [−0.457, −0.339]), indicating that self-reports partially—but not fully—explained age-related performance declines.

In the PC cohort, AI-related self-reports showed qualitatively similar relationships with performance (**Supplementary Fig. S1A–C**). Accuracy was positively correlated with AI-detection confidence (Spearman $\rho = 0.269$, $p < .001$) and more weakly with AI exposure ($\rho = 0.126$, $p < .05$), whereas attitude toward AI showed no reliable association with accuracy ($\rho = 0.033$, $p \geq .05$). A pooled parallel mediation model further indicated a small but significant total indirect effect (ind_total = −0.044, 95% bootstrap CI [−0.089, −0.005]), dominated by the confidence-mediated pathway (ind2 = −0.029, CI [−0.068, −0.003]), while exposure- and

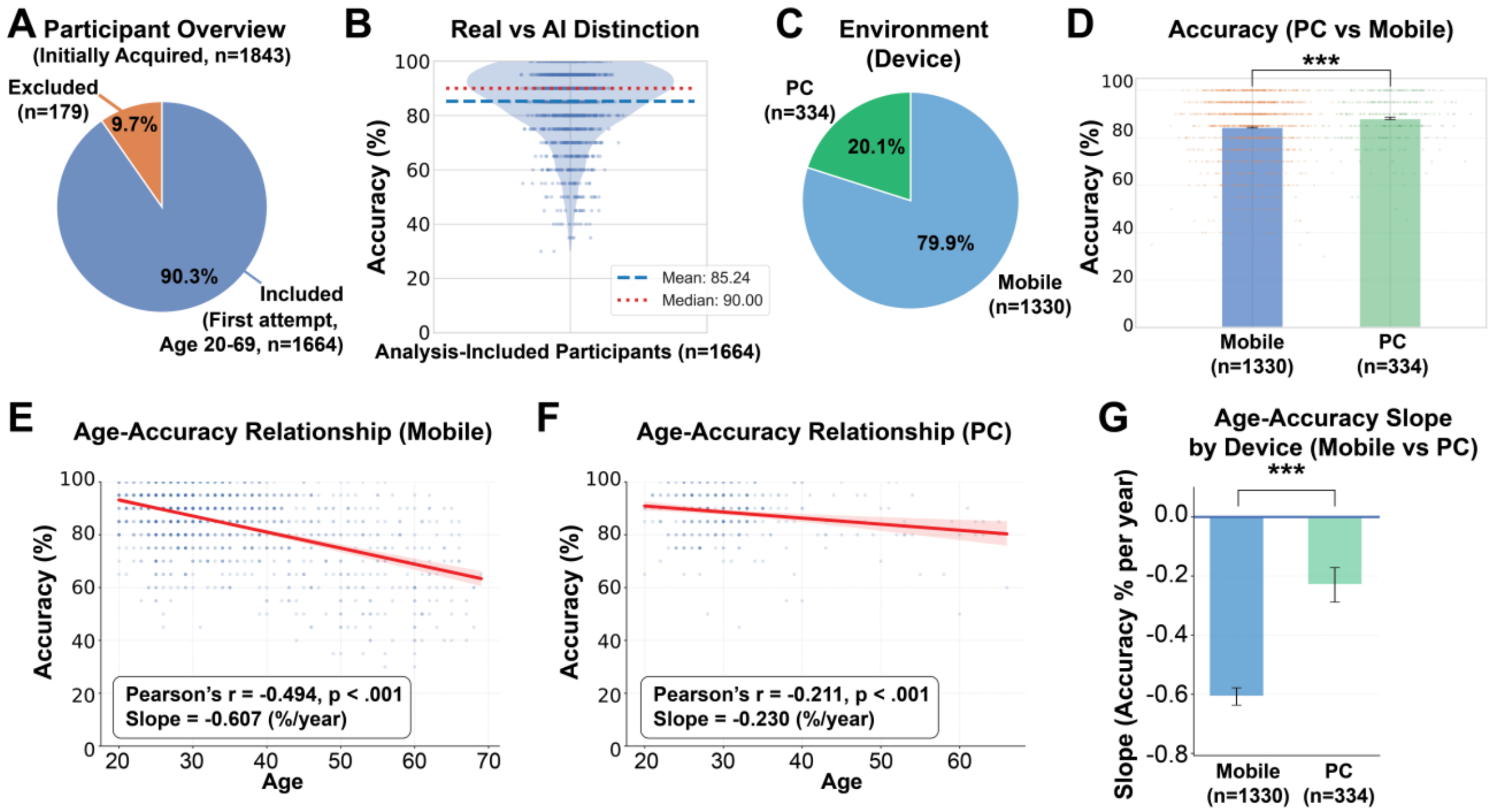

**Figure 2. Sample characteristics and device-dependent age effects on discrimination accuracy**

(A) Participant overview. Of the initially acquired sample (n=1843), a subset was excluded based on pre-specified criteria (e.g., non–first-time participation and age outside 20–69), yielding the final analytic sample (n=1664; first attempt; age 20–69).

(B) Overall accuracy distribution for analysis-included participants, with mean and median indicated.

(C) Device composition within the final analytic sample (web responses are labeled as PC). (D) Accuracy comparison between mobile and PC, shown as mean ± SEM with individual participant values overlaid; PC showed higher accuracy than mobile (Welch's t-test, p<.001).

(E–F) Relationship between age and accuracy within mobile and PC cohorts, with linear regression fits (red line) and summary statistics (Pearson r; p-values shown in-panel), showing a negative age–accuracy relationship in both cohorts.

(G) Estimated age–accuracy slopes by device (bars) with standard errors of slope estimates (error bars) and a slope-difference test, demonstrating a significantly steeper age-related decline in accuracy on mobile than on PC. Asterisks denote statistical significance (* p<.05, ** p<.01, *** p<.001).

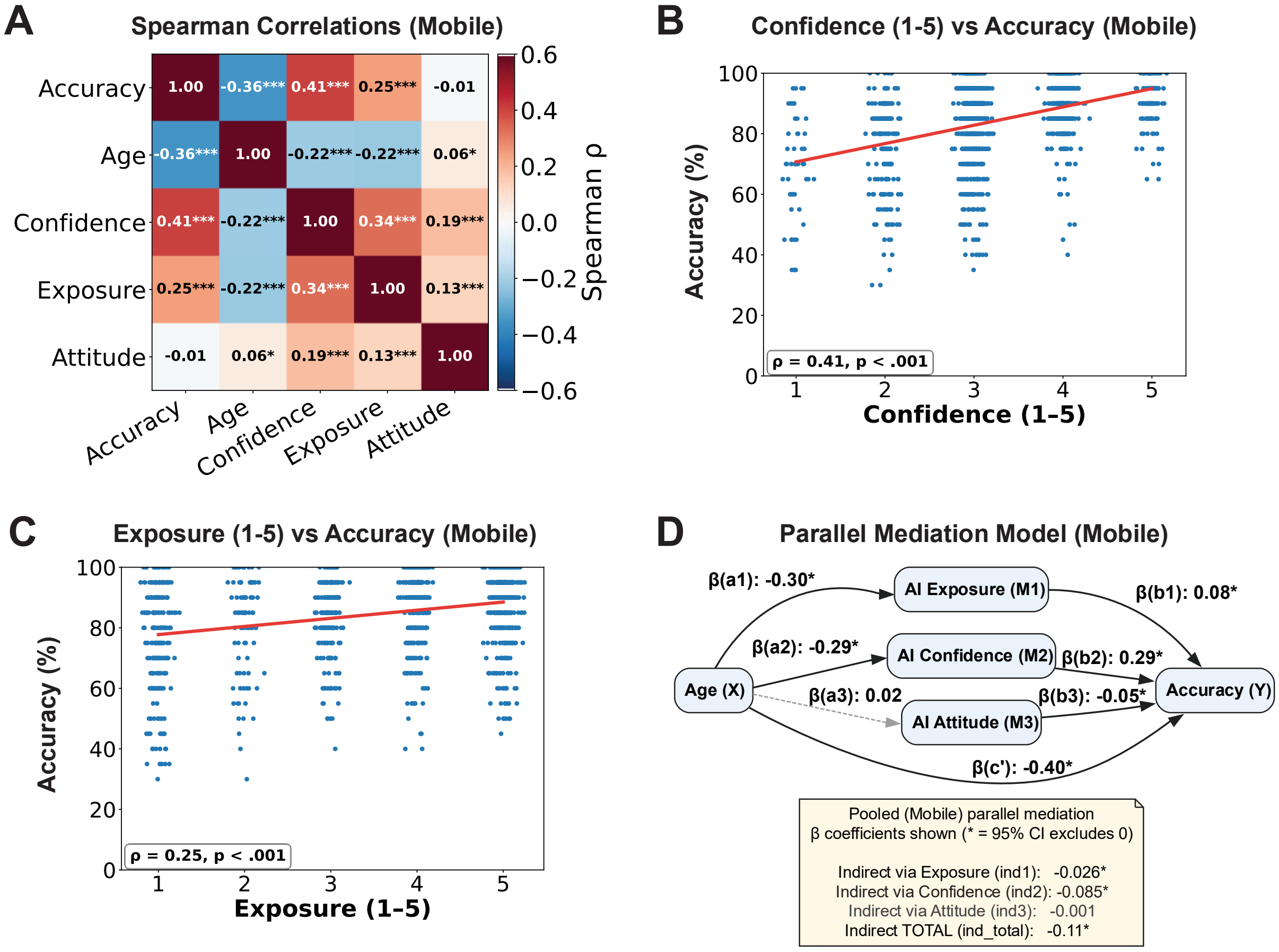

**Figure 3. Accuracy and AI-related factors in the mobile cohort and pooled parallel mediation model**

(A) Accuracy-centered Spearman correlation matrix among age, accuracy, AI exposure, AI-detection confidence, and attitude toward AI in the mobile cohort. Values indicate Spearman ρ; asterisks denote correlation significance (* p < .05, ** p < .01, *** p < .001).

(B–C) Scatter plots showing associations between accuracy and (B) confidence and (C) exposure (mobile cohort). Accuracy is shown on a 0–100% scale; ordinal survey scores are jittered for visualization. Red lines indicate linear fits for visualization, and in-panel annotations report Spearman ρ and p-values.

(D) Pooled parallel mediation model in the mobile cohort. All coefficients are standardized (β; z-scored within the mobile cohort). Path labels follow standard mediation notation: a1–a3 denote Age→mediator paths (Exposure/Confidence/Attitude), b1–b3 denote mediator→Accuracy paths, and c' denotes the direct Age→Accuracy effect controlling for the mediators. Indirect effects are computed as the product of standardized path coefficients for each mediator (β{Age→M} × β{M→Accuracy}; ind1–ind3) and assessed using bootstrap 95% confidence intervals; asterisks in Panel D indicate 95% CIs excluding zero (distinct from correlation p-values in Panel A). Solid arrows denote paths with 95% CIs excluding zero (dashed arrows: n.s.).

attitude-mediated pathways were not reliable (CIs included 0; **Supplementary Fig. S1D**). Notably, interpretation of older-age and sex-stratified effects in PC may be constrained by sparse sampling in the 40s–60s bins (**Supplementary Fig. S0**).

### Sex differences emerge in older ages and are linked to a confidence-mediated pathway in the mobile cohort

Accuracy showed a sex divergence that emerged in later life, with a pronounced drop in females in the 50s–60s (**Fig. 4A**). This pattern was supported by an Age group × Sex interaction in the mobile cohort ($F(4, 1200) = 11.67$, $p < .001$, $\eta^2 = 0.0268$). Post hoc Tukey tests (FWER = 0.05) localized the interaction to older age bins: females underperformed males in the 50s ($\Delta = -7.72$ pp, $p < .05$) and 60s ($\Delta = -16.03$ pp, $p < .001$). Within females, accuracy declined sharply from the 40s to 50s ($\Delta = -12.71$ pp, $p < .001$) and further from the 50s to 60s ($\Delta = -10.28$ pp, $p < .01$). Overall, age group showed a strong main effect ($F(4, 1200) = 123.26$, $p < .001$, $\eta^2 = 0.2831$), whereas the main effect of sex was not significant ($F(1, 1200) = 1.85$, $\eta^2 = 0.0011$).

We next examined whether this late-life divergence was accompanied by sex differences in mediation pathways using sex-stratified parallel mediation and bootstrap-based sex-difference estimates ($\Delta\beta$ = female−male; **Fig. 4B–C**). In **Fig. 4B**, the Age → Confidence path (a2) was negative and significant in both sexes (male a2 = −0.243, 95% bootstrap confidence interval (CI) [−0.326, −0.156]; female a2 = −0.354, CI [−0.420, −0.287]), and the Confidence → Accuracy path (b2) was positive and significant in both sexes (male b2 = 0.239, CI [0.164, 0.318]; female b2 = 0.309, CI [0.245, 0.375]). Importantly, the Age → Confidence path was significantly more negative in females than males (Δa2 = −0.112, CI [−0.218, −0.003]; Δa2 denotes the female−male difference in the Age → Confidence path), whereas the sex difference in the Confidence → Accuracy path was not reliable (Δb2 = 0.070, CI [−0.033, 0.170]; **Fig. 4C**). Consistent with this pattern, the confidence-mediated indirect effect (ind2) was more negative in females than males (male ind2 = −0.058, CI [−0.089, −0.032]; female ind2 = −0.109, CI [−0.142, −0.081]; Δind2 = −0.051, CI [−0.093, −0.009]; **Fig. 4C**), indicating that the sex difference in the confidence-mediated pathway was primarily driven by the Age → Confidence component rather than Confidence → Accuracy. In addition, the direct Age → Accuracy effect (c') was significantly more negative in females than males (male c' = −0.272, CI [−0.352, −0.193] vs female c' = −0.502, CI [−0.582, −0.419]; Δc' = −0.230, CI [−0.345, −0.113]). By contrast, exposure-related mediation effects were of similar magnitude across sexes: the exposure-mediated indirect effect (ind1) was significant within each sex (male ind1 = −0.037, CI [−0.068, −0.010]; female ind1 = −0.021, CI [−0.042, −0.001]) but did not differ reliably by sex (Δind1 = 0.016, CI [−0.018, 0.054]; **Fig. 4C**). Although age-to-attitude

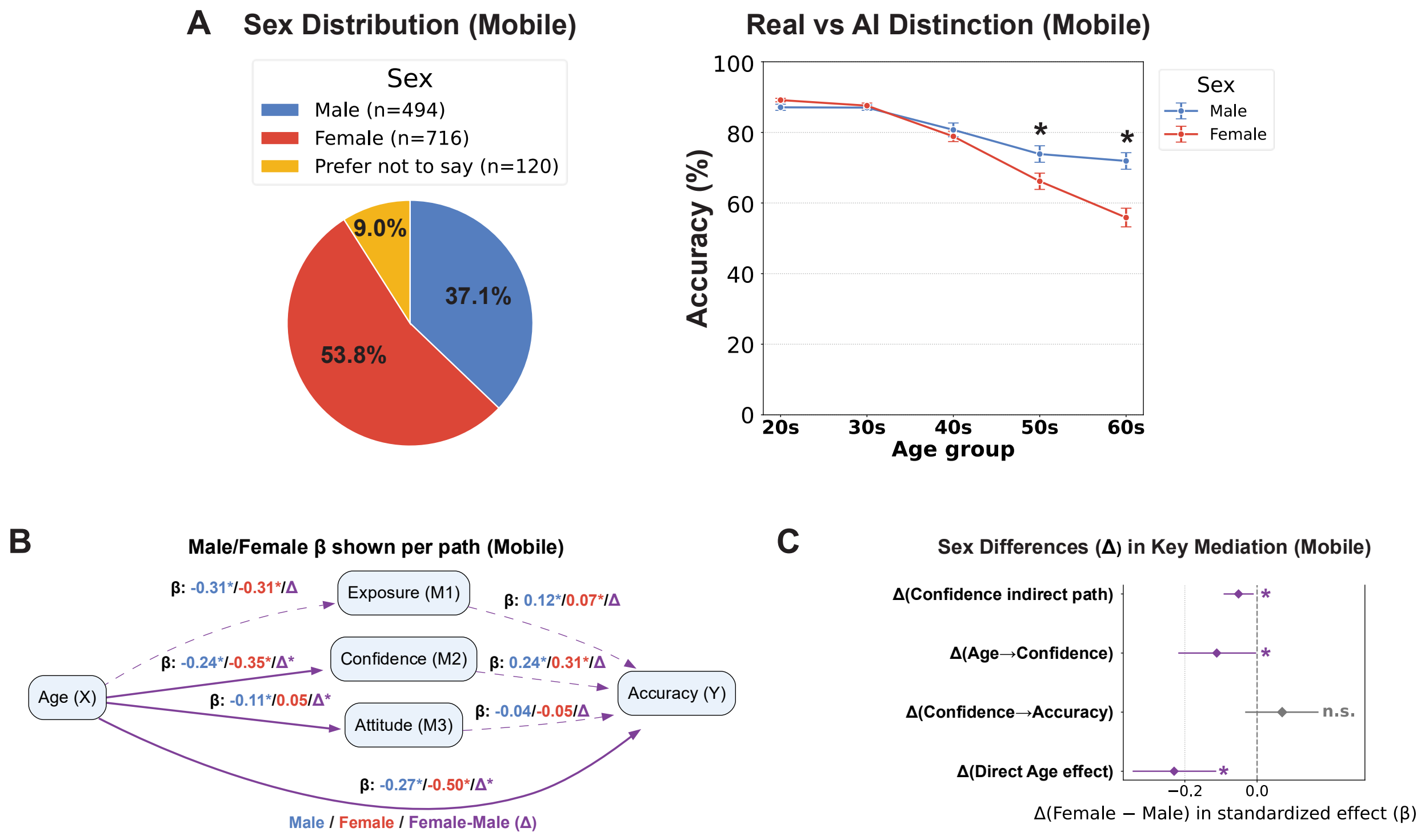

**Figure 4. Sex differences in discrimination accuracy and mediation pathways in the mobile cohort**

(A) Sex distribution (left) and discrimination accuracy across age bins by sex (right; mean ± SEM). A significant Age×Sex interaction was observed ($p < .001$), driven by a sharper performance decline in older females (most pronounced in the 50s–60s bins). Asterisks in Panel A indicate significant female–male differences within the corresponding age bin (post hoc).

(B) Sex-stratified parallel mediation model with standardized coefficients (β). Values on each path show male β / female β / sex difference (female−male, Δ), displayed in blue, red, and purple, respectively. Asterisks attached to male or female coefficients indicate that the corresponding sex-specific path coefficient has a bootstrap 95% CI excluding zero. Δ denotes the sex difference for that path, and Δ* denotes a significant sex difference (bootstrap 95% CI for the difference excludes zero). Purple arrows visualize between-sex differences across paths; solid arrows indicate significant sex differences, whereas dashed arrows indicate non-significant sex differences.

(C) Summary of sex differences in key mediation effects, shown as Δ = female−male (standardized β units) with bootstrap 95% CIs. Asterisks in Panel C indicate that the Δ effect differs from zero (bootstrap 95% CI excludes zero); "n.s." indicates the CI includes zero. The confidence-mediated sex difference is significant for the confidence indirect path and Age→Confidence, but not for Confidence→Accuracy, indicating that the sex difference in the confidence-mediated pathway is primarily driven by the Age→Confidence component. The confidence-mediated indirect effect is computed as β(Age→Confidence) × β(Confidence→Accuracy).

showed a significant sex difference (Δa3 = 0.160, CI [0.059, 0.264]), the Attitude → Accuracy path did not reliably differ from zero in either sex, and the attitude-mediated indirect effect did not show a reliable sex difference (Δind3 = −0.008, CI [−0.020, 0.001]). Together, these findings suggest that age-related reductions in AI-detection confidence are more pronounced in females, contributing to the observed sex divergence in accuracy at older ages in mobile (**Fig. 4A–C**).

In the PC cohort, we did not observe comparable sex-dependent aging effects on accuracy (**Supplementary Fig. S2A**). A two-way ANOVA (Type II) showed a main effect of age group ($F(4, 289) = 3.99$, $p < .01$), with no main effect of sex ($F(1, 289) = 1.64$, $p \geq .05$) and no Age group × Sex interaction ($F(4, 289) = 0.12$, $p \geq .05$). Consistent with this, sex-stratified mediation models did not provide evidence for sex-dependent pathways: bootstrap-based sex-difference estimates (Δβ = female−male) for key mediation effects and path coefficients all had 95% confidence intervals including zero (**Supplementary Fig. S2B–C**). These null sex-difference findings in PC should be interpreted cautiously given the smaller PC sample size and sparse sampling in older age bins (**Supplementary Fig. S0**).

### Decision cues show modest associations with accuracy, but do not explain the sex gap via differential adoption

Participants reported multiple decision cues (strategies) used to classify portraits (**Fig. 5B**), with the most commonly endorsed cues being Painting-like appearance, Texture, and Feeling/Intuition. To test which cues uniquely related to performance beyond co-occurring strategies and demographics, we fit a multivariable ordinary least squares regression controlling for age, sex, and all other strategy indicators (HC3 robust standard errors; FDR correction across strategy terms). In mobile, these high-frequency cues were also among the few that showed robust positive associations with accuracy (**Fig. 5A**), including Texture ($\beta = +4.00$ pp, $q < .001$), Painting-like appearance ($\beta = +2.41$ pp, $q < .05$), and Feeling/Intuition ($\beta = +1.92$ pp, $q < .05$), whereas other commonly reported cues did not show reliable independent associations after adjustment. However, sex differences in the usage of these “effective” cues within age bins did not survive FDR (false discovery rate) correction across age-bin–specific tests (all $q \geq .068$; **Fig. 5C**), suggesting that the observed sex differences in performance are not readily attributable to differential self-reported adoption of these cues alone.

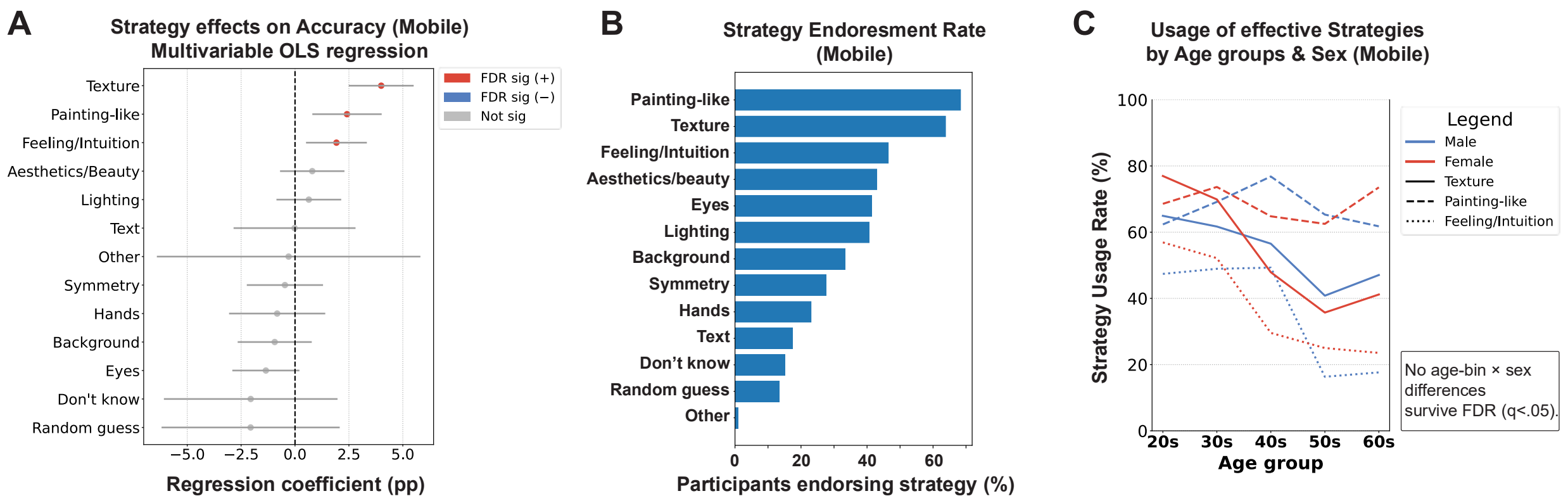

**Figure 5. Decision cues (self-reported strategies) associated with discrimination accuracy in the mobile cohort**

(A) Multivariable ordinary least squares (OLS) regression estimating strategy associations with accuracy while controlling for age, sex, and all other strategy indicators. Points show regression coefficients in percentage points (pp) with 95% confidence intervals computed using heteroskedasticity-consistent (HC3) robust standard errors; colors indicate FDR-corrected significance across strategy terms ($q < .05$).

(B) Decision-cue endorsement rate in the mobile cohort (percentage of participants endorsing each cue; multiple cues could be selected per participant).

(C) Usage rates of the effective cues (Texture, Painting-like, Feeling/Intuition) across age bins and sex. Sex differences in cue usage were tested within each age bin, and none survived FDR correction across age-bin–specific tests (all $q \geq .068$), suggesting that the observed sex differences in performance are not readily attributable to differential self-reported adoption of these cues alone.

In the PC cohort, strategy use patterns were broadly similar, with Texture, Painting-like, and Feeling/Intuition among the most commonly endorsed cues (**Supplementary Fig. S3B**). However, in the multivariable OLS model controlling for age, sex, and all other strategies, no strategy term survived FDR correction in PC (**Supplementary Fig. S3A**). These null strategy–accuracy associations in PC should be interpreted cautiously given the smaller PC sample size (and reduced precision for multivariable estimates), particularly in older age bins (**Supplementary Fig. S0**).

### Unified human-factors model quantifies joint contributions of demographics, self-reports, and strategies

A unified model jointly incorporating demographics (age, sex), AI self-reports (confidence, exposure, attitude), and selected strategy indicators accounted for substantial variance in mobile accuracy (**Fig. 6A**). Standardized coefficients indicated strong negative effects of age ($\beta = -0.396$, $p < .001$) and positive effects of confidence ($\beta = 0.275$, $p < .001$), with smaller contributions from exposure ($\beta = 0.072$, $p < .01$) and selected strategies (Texture $\beta = 0.154$, $p < .01$; Painting-like $\beta = 0.120$, $p < .05$). Feeling/Intuition was not reliably associated with accuracy in the multivariable context ($\beta = -0.017$, $p \geq .05$). Attitude showed a small negative association after adjustment ($\beta = -0.057$, $p < .05$), despite weak bivariate relationships, consistent with a suppression/adjustment effect when controlling for correlated predictors. Nested $R^2$ analyses clarified incremental explanatory contributions (**Fig. 6B**): demographics alone explained substantial variance ($R^2 = 0.262$), adding AI self-reports produced the largest incremental gain ($\Delta R^2 = 0.090$), and strategies provided a smaller additional gain ($\Delta R^2 = 0.009$).

In the PC cohort, the unified model showed smaller overall explained variance, with significant contributions from age, confidence, and Texture, while other self-report and strategy terms were not reliable (S**upplementary Fig. S4A–B**); interpretation is tempered by sparse older-age sampling in PC (**Supplementary Fig. S0**).

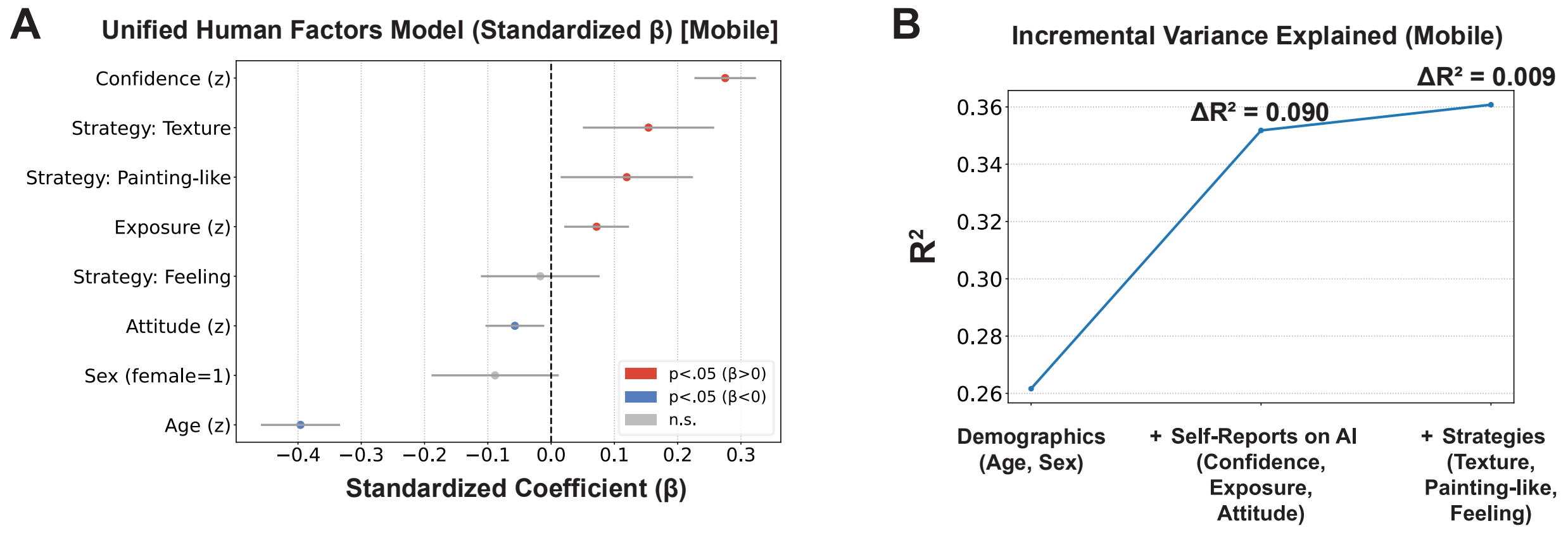

**Figure 6. Unified human-factors model of discrimination accuracy in the mobile cohort**

(A) Standardized regression coefficients (β) from a multivariate model including demographics (age, sex), AI-related self-report measures (confidence, exposure, attitude), and effective strategy indicators (texture, painting-like appearance, feeling/intuition). Points show β with 95% confidence intervals computed using heteroskedasticity-consistent (HC3) robust standard errors (colors indicate $p<.05$ for each coefficient).

(B) Incremental variance explained ($R^2$) in nested OLS models adding predictor blocks sequentially. Demographics alone explained substantial variance in accuracy ($R^2 = 0.262$). Adding AI-related self-reports produced the largest incremental gain ($\Delta R^2 = 0.090$), whereas strategies provided a smaller additional gain ($\Delta R^2 = 0.009$).

### Reaction time shows condition-specific effects beyond overall age-related slowing

Reaction time (RT) increased with age in the mobile cohort (two-way ANOVA main effect of age: $F(4, 1200) = 9.39$, $p < .001$), with no main effect of sex ($F(1, 1200) = 1.12$, $p \geq .05$) and no Age×Sex interaction ($F(4, 1200) = 0.10$, $p \geq .05$; **Fig. 7A**). Trial-level mixed-effects modeling of logRT (participant random intercepts; fixed effects Correctness×Image kind plus age and sex) revealed a robust Correctness×Image-kind interaction ($p < .001$; **Fig. 7B**). Model-based estimated marginal means (back-transformed to seconds) showed a crossover pattern: when responses were correct, RTs were faster for AI than real images (Real–AI = +0.53 s; post hoc Real–AI contrast, $p < .001$), whereas when responses were incorrect, RTs were slower for AI than real images (Real–AI = −0.19 s; $p < .01$; **Fig. 7B**).

In the PC cohort, overall RT also varied by age group ($F(4, 289) = 15.78$, $p < .001$; **Supplementary Fig. S5A**), with no main effect of sex ($F(1, 289) = 0.18$, $p \geq .05$); age-bin estimates remain noisy in older bins due to sparse sampling (**Supplementary Fig. S0**). The trial-level mixed-effects model again showed a significant Correctness×Image-kind interaction ($p < .001$; **Supplementary Fig. S5B**). However, unlike mobile, the post hoc Real–AI contrast was reliable only for correct responses (AI faster than real; $p < .001$) and not for incorrect responses ($p \geq .05$), indicating partial replication of the condition-specific RT pattern in PC (**Supplementary Fig. S5B**).

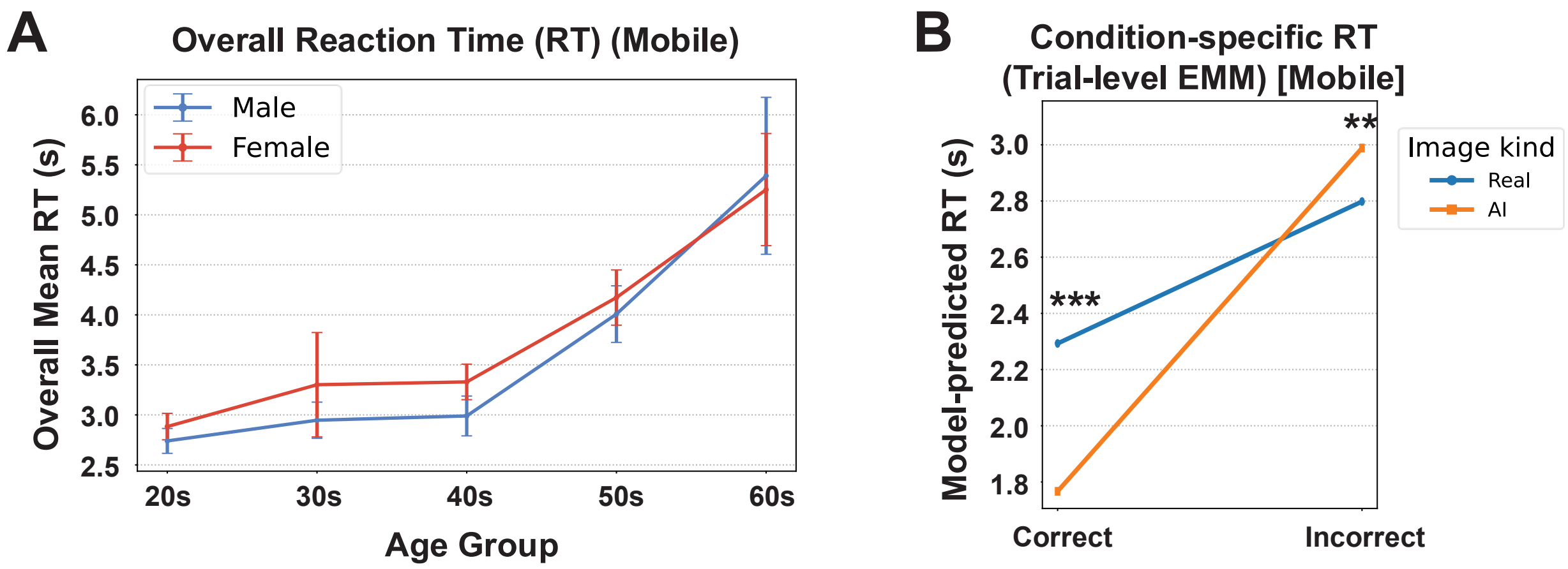

**Figure 7. Reaction-time (RT) evidence for condition-specific verification demands**

(A) Overall mean RT by age group and sex in the mobile cohort (mean ± SEM; RT in seconds). Two-way ANOVA showed a significant main effect of age on RT ($p < .001$), with no main effect of sex ($p \geq .05$) and no Age×Sex interaction ($p \geq .05$).

(B) Model-based estimated marginal means (EMMs) from a trial-level linear mixed-effects model (LMM) of logRT with participant random intercepts. The model included fixed effects of Correctness × Image kind (Real vs AI), age, and sex, and revealed a significant Correctness×Image-kind interaction ($p < .001$), indicating that the RT difference between real and AI images depends on whether responses are correct or incorrect. EMMs were back-transformed to seconds for visualization. Asterisks indicate post hoc Real–AI contrasts within each correctness condition (** $p < .01$, *** $p < .001$).

### Generator-dependent difficulty: performance differs for ChatGPT-4o vs Imagen 3 (via Gemini 2.5)

Finally, we tested whether discrimination performance depended on the generator used to create AI portraits (stimuli created July 2025; **Fig. 8**). In mobile ($N = 1330$), accuracy was lower for ChatGPT-4o than for Imagen 3 (79.86 ± 0.62% vs 88.33 ± 0.48%; paired t-test, $t(1329) = -13.35$, $p < .001$; **Fig. 8A**), and responses were slower for ChatGPT-4o (3.205 ± 0.145 s vs 2.677 ± 0.087 s; paired t-test, $t(1329) = 4.13$, $p < .001$; **Fig. 8B**). Age-related decline was also steeper for ChatGPT-4o than for Imagen 3 (−0.756 vs −0.479 pp/year; age×generator interaction with HC3 robust standard errors, $p < .001$; **Fig. 8C–D**). The PC cohort showed the same qualitative pattern (**Supplementary Fig. S6**): accuracy was lower for ChatGPT-4o (83.11 ± 1.15% vs 91.68 ± 0.78%; paired t-test, $t(333) = -6.69$, $p < .001$) and RT was longer for ChatGPT-4o (3.529 ± 0.297 s vs 2.933 ± 0.219 s; paired t-test, $t(333) = 2.19$, $p < .05$), with generator-dependent differences in age-related decline ($p < .01$), though interpretation for older ages is constrained by sparse sampling in older PC bins (**Supplementary Fig. S0**).

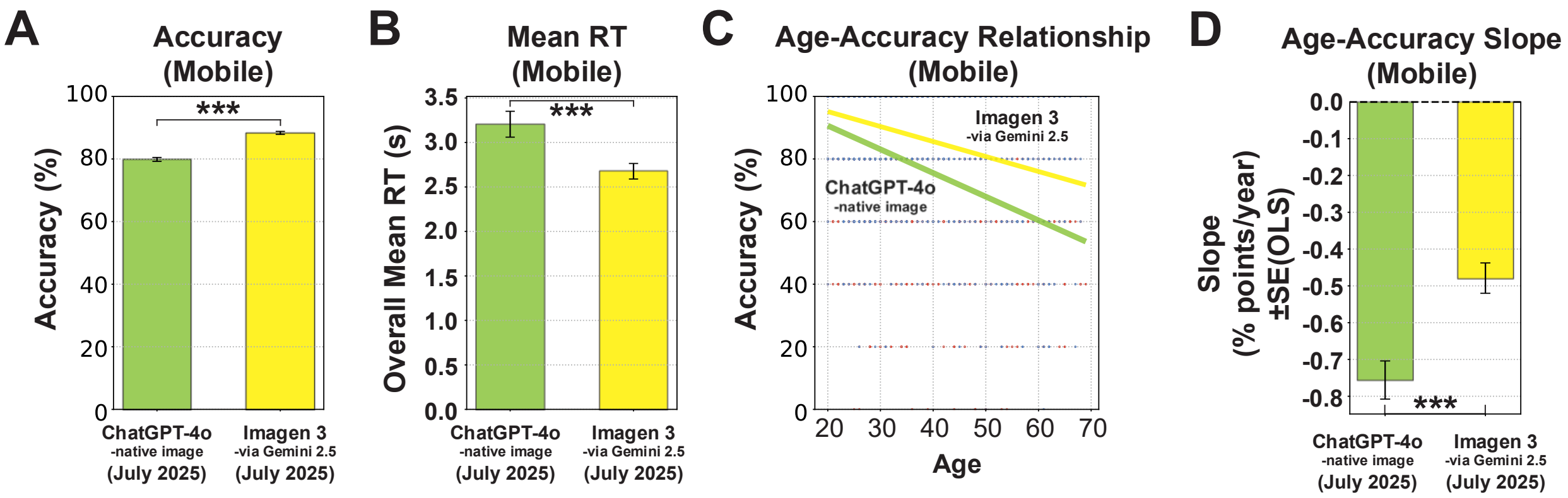

**Figure 8. Discrimination performance depends on the AI portrait generator used to create the stimuli (stimuli created July 2025)**

(A) Participant-level accuracy for AI portraits generated by ChatGPT-4o (native images) versus Imagen 3 (via Gemini 2.5) in the mobile cohort; accuracy was lower for ChatGPT-4o (79.86% vs 88.33%; paired t-test, $p < .001$).

(B) Participant-level mean reaction time (RT, seconds) was longer for ChatGPT-4o (3.205 s vs 2.677 s; paired t-test, $p < .001$).

(C) Age–accuracy relationships plotted separately by generator with fitted linear trends.

(D) Estimated age–accuracy slopes (percentage points/year) differed by generator (−0.756 vs −0.479 %/year); slope differences were tested using an age×generator interaction model with heteroskedasticity-consistent (HC3) robust standard errors ($p < .001$).

## Discussion

This study provides a time-stamped snapshot of human performance in discriminating real face photographs from AI-generated portraits produced by models available as of July 2025. Mean accuracy was high, but that overall value concealed substantial heterogeneity across device context, age, sex, self-reported confidence and exposure, and, to a lesser extent, self-reported decision cues and response dynamics.

At first glance, the relatively high mean accuracy may seem difficult to reconcile with prior work reporting modest or near-chance human detection of deepfakes and other synthetic media, often accompanied by a tendency to accept content as authentic[1–4]. The present design, however, does not demonstrate that human detection ability has improved over time. A narrower and more direct conclusion is that, under this particular July 2025 stimulus set and single-interval REAL/AI task, average performance was high. One possible interpretation is that growing exposure to AI-generated media may have contributed to partial adaptation, but that claim remains inferential in the absence of longitudinal or repeated cross-sectional data. Recent work showing that targeted training can improve synthetic-face detection makes such an interpretation plausible, though not directly established here[8,9]. More generally, the present findings reinforce that detectability is strongly dependent on the generator, stimulus pool, and decision context. Our results should therefore be interpreted as a time-stamped capability snapshot rather than a universal benchmark.

Against that backdrop, accuracy declined with age in both device cohorts, and the age-related decline was steeper on mobile than on PC. This pattern suggests that age-related differences in real-world AI-portrait detection are shaped not only by participant characteristics, but also by viewing context. Smaller displays, more variable environments, and other uncontrolled aspects of mobile use may amplify existing difficulty, although those specific mechanisms were not directly measured here. AI-detection confidence and AI exposure were positively associated with accuracy and statistically accounted for part of the age–accuracy association, with the confidence pathway especially prominent in the mobile cohort and present, though smaller, in the PC cohort. Because a substantial direct age effect remained after the mediators were included, these self-reported factors should be regarded as partial rather than complete explanations. One plausible interpretation is that confidence indexes subjective certainty or metacognitive calibration when evaluating contemporary synthetic portraits. Older participants may therefore have experienced greater uncertainty, particularly under mobile viewing conditions, and that uncertainty may have contributed to lower accuracy.

A further result was a late-life sex divergence in the mobile cohort. Female participants showed lower accuracy in the 50s and 60s, and sex-stratified mediation analyses suggested that confidence contributed to that pattern. Importantly, there was no clear evidence that this sex difference was driven by a stronger Confidence→Accuracy link in females. Rather, the confidence-mediated difference was primarily driven by a more negative Age→Confidence association in females, while a more negative direct age effect remained after accounting for the mediators. Confidence should therefore be interpreted as a contributing pathway rather than a complete explanation of the female disadvantage at older ages. More broadly, the result suggests that age-related changes in confidence may not be uniform across sexes in this task context. Future work should examine whether this pattern reflects differential exposure to AI media, differential adoption of AI tools, broader technology self-efficacy, or other unmeasured contextual variables.

Self-reported decision cues showed modest but interpretable associations with performance. In the mobile cohort, texture and painting-like appearance were the most consistent positive predictors across the strategy-specific and unified models, whereas feeling/intuition was less robust once other predictors were entered jointly. At the same time, sex differences in endorsement of these cues did not survive multiple-testing correction. This makes it unlikely that the late-life sex divergence can be explained simply by differential self-reported strategy adoption. Instead, these findings suggest that cue labels capture only part of the underlying process: participants may endorse similar cue categories while attending to different concrete image properties, and successful discrimination may depend more on extracting and weighting subtle evidence than on explicitly naming a strategy.

The reaction-time analyses provide complementary process-level information. In the mobile cohort, correct AI judgments were faster than correct real judgments, whereas incorrect AI judgments were slower than incorrect real judgments. The same interaction was also significant in the PC cohort, but the incorrect-trial simple effect did not replicate there. One possible interpretation is that some AI portraits contained sufficiently salient cues to permit rapid rejection, whereas more ambiguous synthetic portraits elicited longer evaluation and still produced errors. By contrast, real portraits may have required more verification when correctly accepted, but could be rejected relatively quickly when misclassified as AI. On this reading, judgment failures are not merely weaker versions of successful judgments, but may involve qualitatively different decision dynamics.

Generator family also mattered. Within the present stimulus set and mirroring pipeline, ChatGPT-4o portraits were harder to classify than Imagen 3 portraits, elicited slower responses, and showed steeper age-related decline in both cohorts, with the clearest pattern in the larger mobile sample. This result underscores that the detectability of AI-generated portraits is not a single fixed property of “AI images,” but depends on the generator and on the prompting and task conditions under which stimuli are produced and judged.

One possible interpretation is that the ChatGPT-4o portraits more often matched participants' photographic expectations while offering fewer overtly diagnostic deviations, thereby increasing ambiguity, particularly for older participants and under mobile viewing conditions. That interpretation should remain specific to the present stimulus pool rather than generalized as a stable ranking of models.

Several limitations should be considered when interpreting these findings. First, this study provides a time-bounded snapshot: the stimuli reflect models available as of July 2025, and both generator capabilities and public familiarity with their outputs may shift rapidly. Second, only two generator families and a fixed portrait set were tested, so the results may not generalize to other models, prompting styles, portrait subtypes, or post-processing pipelines. Third, recruitment was public, online, and gamified, yielding a self-selected sample and a strongly mobile-skewed device distribution. Viewing conditions were uncontrolled, including screen size, brightness, distance, ambient light, and distraction. Fourth, the single-image REAL/AI task with balanced trial counts differs from many real-world settings, where base rates are unknown and judgments are embedded in richer social context. Fifth, exposure and confidence were self-reported and ordinal, and the strategy checklist captured endorsed cue categories rather than the fidelity with which those cues were actually applied during viewing. Finally, some subgroup estimates, especially in older PC bins, were based on sparse data and should therefore be interpreted cautiously.

Despite these limitations, the overall pattern is clear. AI-portrait detection in this dataset was neither uniformly poor nor uniformly strong. Instead, it depended systematically on who was judging, on what device, and on which generator family produced the image. That combination of demographic, contextual, and generator-specific structure is consistent with treating AI-portrait detection as a human-factors problem rather than as a single fixed level of human performance.

## Conclusion

Using a large-scale web-based experiment, we provide a time-stamped snapshot of human discrimination between real face photographs and AI-generated portraits produced by models available as of July 2025. Average accuracy was high, but performance was not uniform. Accuracy declined with age in both device cohorts and more steeply on mobile than on PC. In the mobile cohort, a late-life female disadvantage emerged, to which confidence-related pathways contributed but did not fully account for the pattern. Within the present stimulus set, generator family also mattered: ChatGPT-4o portraits were harder to classify than Imagen 3 portraits and elicited slower responses. Together, these findings support treating AI-portrait detection as a time-sensitive human-factors problem shaped by age, sex, device context, confidence, and generator family within a rapidly changing technological landscape.

## Methods

### 1. Ethics and data-collection period

This study was approved by the Institutional Review Board (IRB) at Hwasung Medi-Science University (Approval No. HSMUIRB-2025-06). Participation was voluntary and obtained via digital informed consent. Data were collected from August 2025 through January 2026. No identifying information (e.g., email address) was collected, and no monetary or material compensation was provided. To encourage engagement without financial incentives, participants received an in-app score report and a non-monetary "title" with an illustrative reward image at the end of the task (see below).

### 2. Stimuli

A total of 210 face-portrait stimuli (512×512 px) were prepared (**Fig. 1A**; see Data and Materials Availability section).

- **Real portraits (n=70):** Selected from the Flickr-Faces-HQ (FFHQ) dataset[10] to ensure broad variation in appearance.
- **AI portraits (n=140):** Generated from the same 70 real portraits using two generators:
    - **ChatGPT-4o native image**: 70 images
    - **Imagen 3 (via Gemini 2.5)**: 70 images
      Thus, the final stimulus pool was 70 Real + 70 ChatGPT-4o + 70 Imagen 3 (via Gemini 2.5) = 210.

#### "Mirroring" via reverse prompting (model-matched generation)

To create high-fidelity AI counterparts while minimizing prompt-engineering bias, we used a *mirroring* procedure. Each generator received the same real portrait and was instructed to write a prompt to recreate the image as accurately as possible. The resulting prompt was then used **within the same generator** to produce the final AI portrait (i.e., prompt-authoring and image generation were both performed by the same model family).

#### Standardization / masking

To reduce the possibility that participants relied on trivial corner artifacts or other consistent UI cues, a small white rectangular patch was overlaid on the bottom-right corner of **all** images (real and AI) before deployment.

## 3. Web-based task and procedure

The experiment was delivered as a web application (HTML/JavaScript) with Google Firebase/Firestore used for real-time logging of trial responses and survey data (**Fig. 1B**; see Data and Materials Availability section).

### 3.1 Language and device selection

Upon entry, participants first selected **Mobile** or **PC (Web)**. The interface language (**Korean or English**) was chosen after the practice block; practice screens and their feedback were therefore shown bilingually (Korean and English). The selected language determined the presentation language for task instructions and survey items, whereas the response buttons were labeled “REAL” and “AI” in both language versions; the stimuli, task structure, and sampling procedure were otherwise identical across language versions. The device selection was recorded as a self-reported label and used to enable device-stratified analyses; the displayed image size was determined by the browser viewport width rather than by this selection.

### 3.2 Practice trials with feedback

Participants first completed 4 practice trials (2 real, 1 ChatGPT-4o AI, 1 Imagen 3 AI), presented before the language choice and therefore shown with bilingual (Korean/English) text. During practice, immediate feedback was shown (correct/incorrect), ensuring participants understood the task.

### 3.3 Main trials without feedback (to prevent contamination)

Participants then completed 20 main trials with no trial-by-trial correctness feedback. This design prevented learning/contamination from explicit label exposure during the main data-collection phase.

### 3.4 Trial timing and response

Each trial began with a central fixation cross (“+”) displayed for 1000 ms, followed by the portrait image. Participants classified the single portrait as “REAL” or “AI”. Each trial presented one image only; “REAL” and “AI” are the two response options, not two simultaneously presented stimulus alternatives. Responses were not time-limited. Reaction time (RT) was recorded from stimulus onset to button click.

### 3.5 Random stimulus sampling per session

For each participant/session, the app randomly sampled 10 real + 10 AI images from the 210-image pool, with AI trials balanced to include five ChatGPT-4o and five Imagen 3 (via Gemini 2.5) images. Any identity shown during practice was excluded from that session's main trials in all three of its versions (real, ChatGPT-4o, and Imagen 3), and no identity was repeated within the main trials.

### 3.6 Intermission and fatigue control

Between trials, an intermission screen was presented and participants could proceed at their own pace, reducing cumulative fatigue and allowing self-paced participation.

## 4. Recruitment and gamification

Recruitment used a snowball approach via online communities and social-media sharing. The study incorporated non-monetary gamification to encourage completion and organic sharing: after finishing the task, participants received a performance-based "perceptual title" (one of eight categories) accompanied by a character-style reward image. This reward was presented only after completion and did not alter task incentives during the trials. (No monetary/material rewards were provided.)

## 5. Post-task survey measures

After the 20 main trials, participants completed a brief survey.

**Included in the paper (core variables):**

- **Demographics**
  - **Age** (mandatory; analyses focus on adults; see filtering below)
  - **Sex** (mandatory): male / female / prefer-not-to-say
- **AI-related self-report measures**
  - **AI exposure frequency** (5-point ordinal): Never(1) ... Daily(5)

- **AI detection confidence** (5-point ordinal): Very not confident(1) … Very confident(5)
- **Attitude toward AI-generated images/videos** (mapped to a centered scale): Very negative(−2), Negative(−1), Neutral(0), Positive(+1), Very positive(+2)

- **Self-reported decision cues / strategies**

  Participants selected any that applied from a fixed checklist of cues (e.g., hands/fingers, facial details, background inconsistency, texture, lighting, symmetry, overall “feeling/intuition,” etc.), with an optional free-text “other” field.

**Not emphasized in the paper (collected but omitted for brevity unless needed):** Nationality, occupation, MBTI, detailed AI-tool lists, and other optional survey fields can be left out of the main Methods and reserved for supplementary description only if reviewers request it.

**6. Data preprocessing and final analytic sample**

A multi-stage filtering pipeline was used.

1. **Initial collection:** 1,843 participant records were logged.
2. **Repeat-participation filter:** Participants self-reported whether it was their first attempt (“firstTime”). Analyses retained only first-time attempts to reduce learning effects from repeated play.
3. **Adult and plausibility filter:** Participants outside the 20–69 age range were excluded. This served (i) an ethical constraint (adult-only analysis) and (ii) quality control to remove implausible entries.
4. **Final analytic sample:** 1,664 participants remained (Mobile n=1,330; PC n=334). Sex categories included male/female/prefer-not-to-say in descriptive summaries; inferential sex-stratified analyses used male/female unless otherwise stated. Note: the public de-identified release of these data yields N = 1,667 under the same two filters (first attempt only; ages 20–69), a three-participant difference from the N = 1,664 reported here. The repository attributes this to an independently reconstructed de-identified export and records the exact cause as unresolved; overall accuracy differs by 0.06 percentage points (85.24% vs 85.18%), which does not affect any conclusion.

## 7. Data analysis and statistics

All analyses were performed in Python 3 using standard scientific computing libraries. Accuracy was defined as the percentage of correct classifications across the 20 main trials for each participant. Reaction time (RT) was defined as the elapsed time from stimulus onset to response selection; RT is reported in seconds for visualization, and log-transformed RT (logRT) was used for trial-level modeling.

### 7.1 Group comparisons and age effects

Device effects (mobile vs PC) and generator effects [ChatGPT-4o vs Imagen 3 (via Gemini 2.5)] on participant-level accuracy and mean RT were tested using Welch's t-test (between-device comparisons) or paired t-tests (within-participant generator comparisons), with nonparametric Wilcoxon signed-rank tests used as a robustness check when applicable. Age–accuracy (and age–RT) trends were summarized with linear regression slopes; slope differences between groups (e.g., device or generator) were tested using interaction models (e.g., age × device; age × generator). For visualization, slope standard errors were obtained from ordinary least squares fits within each group.

### 7.2 Correlation analyses

Associations among accuracy, age, and AI-related self-report measures were summarized using correlation analyses. Because several survey variables were ordinal (e.g., exposure frequency and confidence ratings), Spearman rank correlations ($\rho$) were used as the primary measure in correlation matrices and bivariate summaries. Pearson correlations ($r$) were additionally reported for relationships treated as approximately linear/continuous (e.g., age–accuracy summaries accompanying linear regression fits).

### 7.3 Mediation analyses

To examine whether AI-related self-report measures statistically accounted for age-related differences in accuracy, we fit a parallel mediation model with age as the predictor (X), accuracy as the outcome (Y), and AI exposure, AI-detection confidence, and AI attitude as parallel mediators. Path coefficients were reported as standardized coefficients ($\beta$; z-scored within cohort). Indirect effects were computed as the product of the standardized path coefficients for each mediator ($a \times b$) and assessed using bootstrap 95% confidence intervals (significance defined as CI excluding zero). Sex-stratified mediation models and bootstrap-based sex-difference estimates ($\Delta\beta$ = female−male) were used to characterize sex-dependent pathways in the mobile cohort.

### 7.4 Strategy (decision-cue) analyses

Self-reported decision cues (strategies) were analyzed in two complementary ways. First, multivariable ordinary least squares (OLS) regression models were fit to estimate unique strategy associations with accuracy while controlling for age, sex, and all other strategy indicators. Heteroskedasticity-consistent (HC3) robust standard errors were used for confidence intervals, and p-values for strategy coefficients were corrected for multiple testing using the Benjamini–Hochberg false discovery rate (FDR) procedure and reported as q-values. Second, for age-bin-specific usage comparisons, sex differences in cue endorsement were tested within each age bin using 2×2 $\chi^2$ tests, with Benjamini–Hochberg FDR correction applied across the resulting set of age-bin–specific tests. Caveat: 202 participants (12.1%) have every checklist item except the free-text option marked, together with a stray token from the form's select-all control — a serialization fault rather than genuine endorsement. Excluding them lowers the rates of the affected items without changing their ordering (e.g., painting-like 68.6% to 64.2%, texture 65.3% to 60.5%, eyes 42.8% to 34.9%), so the strategy endorsement rates should be read as approximate.

### 7.5 Unified human-factors model and incremental variance

A unified multivariable model was used to jointly evaluate demographics (age, sex), AI-related self-report measures (exposure, confidence, attitude), and selected strategy indicators as predictors of accuracy. Standardized regression coefficients (β) were reported with HC3 robust standard errors. Nested-model $R^2$ analyses were used to quantify incremental variance explained by sequentially adding predictor blocks (demographics → self-reports → strategies).

### 7.6 Trial-level Reaction Time (RT) modeling

To test condition-specific verification demands, we analyzed trial-level RTs using a linear mixed-effects model (LMM) on logRT with participant random intercepts. Fixed effects included Correctness (correct vs incorrect), Image kind (real vs AI), their interaction, and covariates age and sex. For visualization, model-based estimated marginal means (EMMs) were back-transformed to seconds. We tested condition-specific differences using Real–AI contrasts within each correctness level (simple effects). Contrast p-values were obtained from Wald tests of linear contrasts of the fixed effects.

### 7.7 Significance threshold and reporting conventions

Unless otherwise specified, statistical tests used a two-sided significance threshold of $\alpha = 0.05$. For figures, asterisks denote significance as $p < .05$ (*), $p < .01$ (**), and $p < .001$ (***). For bootstrap-based mediation results, significance was defined by the bootstrap 95% confidence interval excluding zero (rather than a p-value). Where multiple hypotheses were tested (e.g., across strategy terms), false discovery rate (FDR; Benjamini–Hochberg) correction was applied and reported as q-values.

## Data availability

The de-identified trial-level dataset, stimulus metadata and annotations, and the AI-generated portraits used in this study are available in the project repository. FFHQ image-level attribution, license information, and modification notes are provided in Supplementary Table S1 as a version-pinned CSV file in the repository. The public experiment web application is also available online. Additional de-identified fields and raw logs are available from the corresponding author upon reasonable request.

## FFHQ attribution and licensing note

Real photographs were sampled from the Flickr-Faces-HQ (FFHQ) dataset[10]. According to the FFHQ documentation, the dataset metadata, scripts, and documentation are distributed under CC BY-NC-SA 4.0, whereas individual images retain their original Flickr licenses and attribution metadata. Reuse of FFHQ-derived materials therefore remains subject to the original FFHQ terms and the license conditions of the corresponding Flickr images. Image-level attribution, license information, and modification notes for FFHQ-derived stimuli are provided in Supplementary Table S1.

**Acknowledgements**

This work was supported by the HSMU Research Grant (2025).

## Author contributions

S.W.K. conceived and designed the study, generated the stimuli, deployed the experiment, collected the data, performed the analyses, and wrote the manuscript. S.Y.K. contributed to study design and data analysis. Both authors reviewed and approved the final manuscript.

## Competing interests

The authors declare no competing interests.

## Supplementary Figures

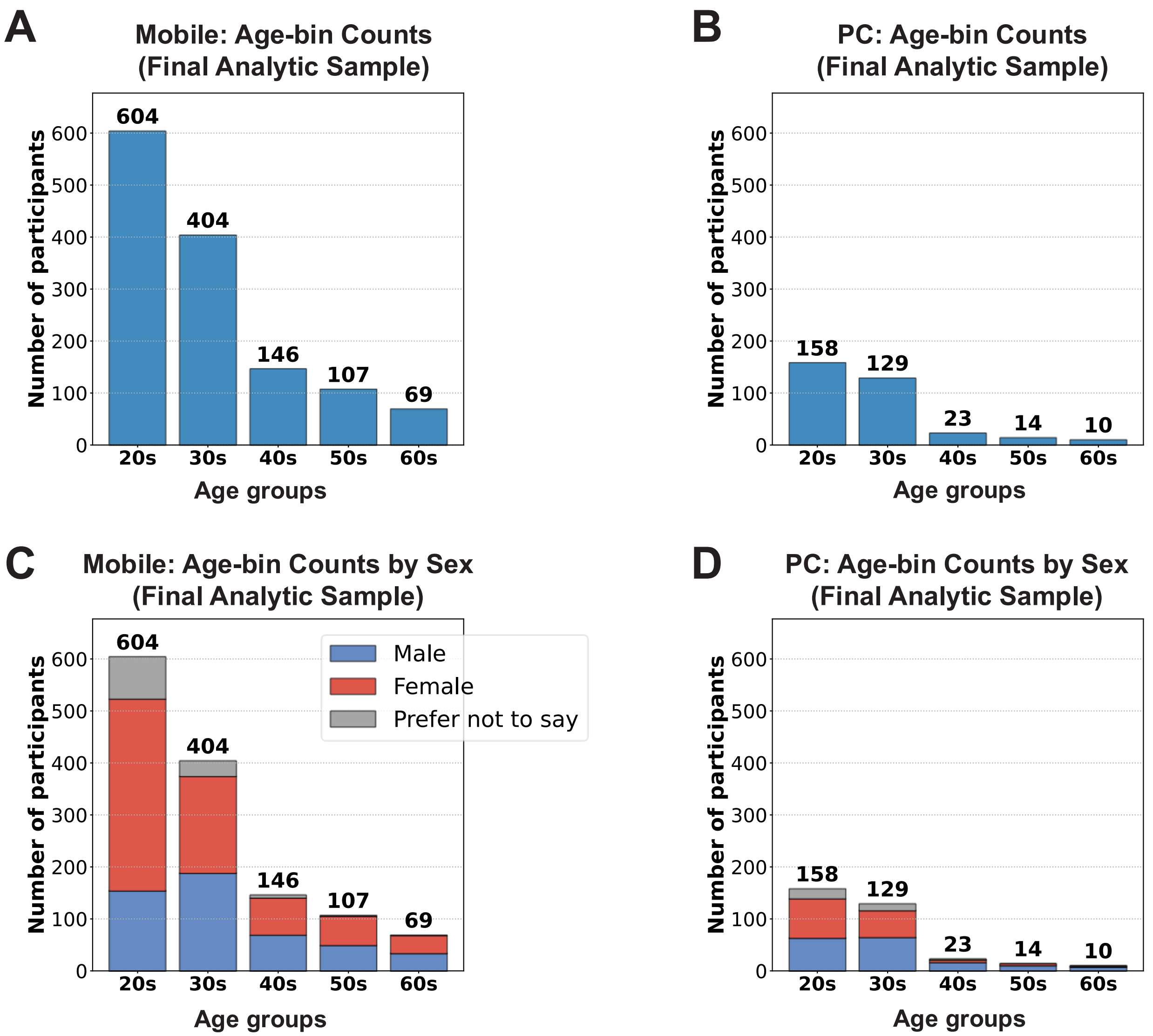

**Supplementary Figure S0. Participant age distributions by device and sex in the final analytic sample**

(A–B) Counts of participants within each age bin (20s–60s) for the mobile and PC cohorts after applying the final inclusion criteria (first attempt; age 20–69).

(C–D) The same age-bin counts stratified by sex (male, female, prefer not to say), shown as stacked counts. Notably, the PC cohort includes relatively few participants in older age bins (40s–60s), which limits statistical power for sex-stratified analyses in that cohort.

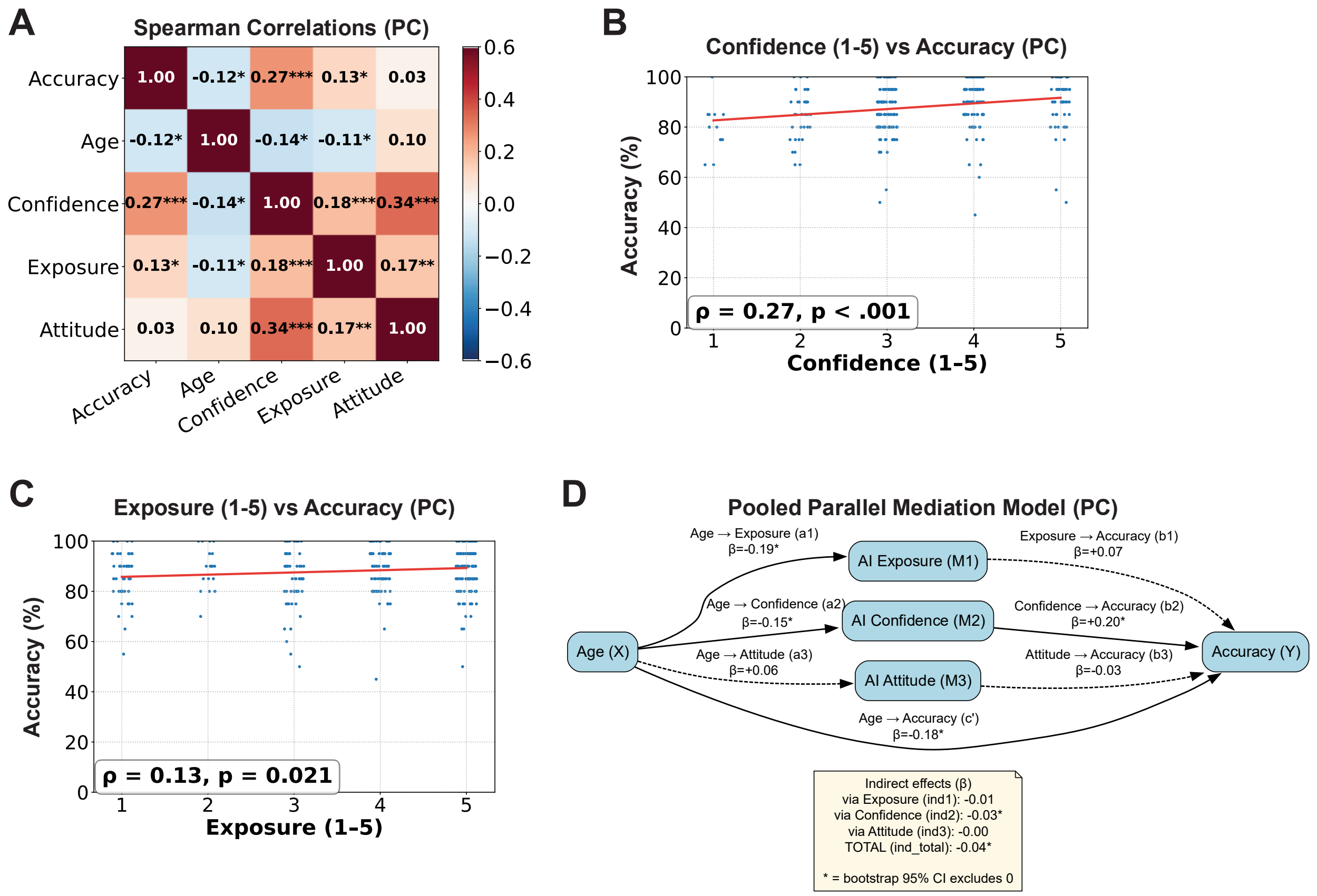

**Supplementary Figure S1. AI-related self-reports and parallel mediation model in the PC cohort**

(A) Accuracy-centered Spearman correlation matrix among accuracy, age, AI-detection confidence, AI exposure frequency, and attitude toward AI in the PC cohort. Values indicate Spearman ρ; asterisks denote correlation significance (* $p < .05$, ** $p < .01$, *** $p < .001$). (B–C) Associations between accuracy and (B) confidence and (C) exposure in the PC cohort. Accuracy is shown on a 0–100% scale; ordinal self-report scores are jittered for visualization. Red lines indicate linear fits for visualization, and in-panel annotations report Spearman ρ and p-values.

(D) Pooled parallel mediation model (standardized coefficients, β) relating age to accuracy through exposure, confidence, and attitude in the PC cohort. Path labels follow standard mediation notation: a1–a3 denote Age→mediator paths, b1–b3 denote mediator→Accuracy paths, and c' denotes the direct Age→Accuracy effect controlling for the mediators. Solid arrows indicate paths with bootstrap 95% confidence intervals excluding zero (dashed arrows: n.s.). Indirect effects are computed as products of standardized path coefficients (ind1–ind3) and are marked with * when the bootstrap 95% CI excludes zero. In the PC cohort, the confidence pathway remains the most prominent indirect route among the three mediators.

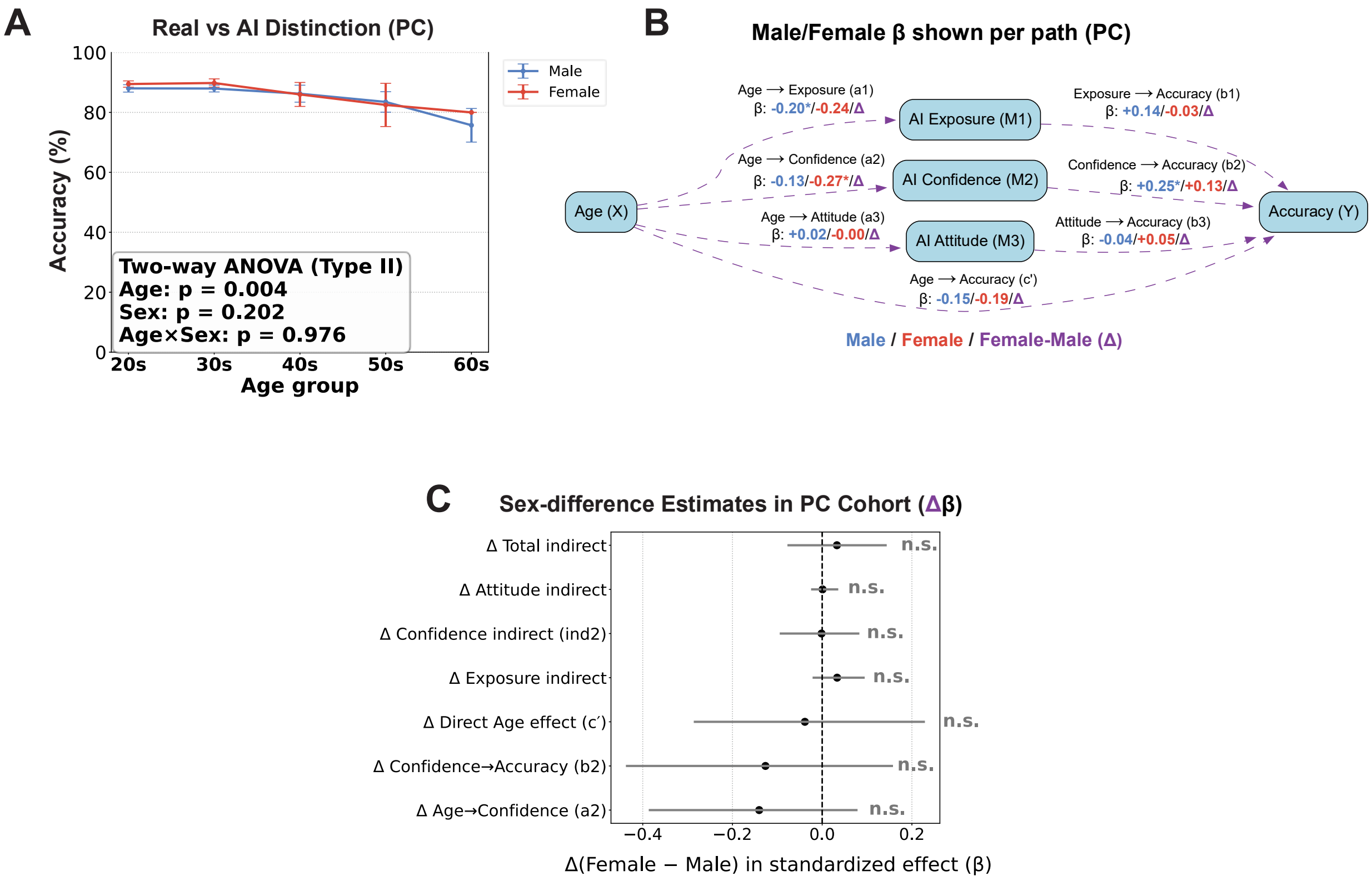

**Supplementary Figure S2. Sex effects on discrimination accuracy and mediation pathways in the PC cohort**

(A) Discrimination accuracy across age bins by sex in the PC cohort (mean ± SEM). A two-way ANOVA (Type II) showed a significant main effect of age group ($p < 0.01$), with no main effect of sex ($p \geq .05$) and no Age×Sex interaction ($p \geq .05$).

(B) Sex-stratified parallel mediation model in the PC cohort (standardized coefficients, β) relating age to accuracy through exposure, confidence, and attitude. Values on each path show male β / female β / sex difference (female−male, Δ) for the corresponding path (a1–a3: Age→mediator; b1–b3: mediator→Accuracy; c': direct Age→Accuracy effect controlling for mediators), displayed in blue, red, and purple, respectively. Purple arrows visualize path-level sex differences; solid arrows indicate significant sex differences, whereas dashed arrows indicate non-significant sex differences. Asterisks attached to male or female coefficients indicate that the corresponding sex-specific path coefficient has a bootstrap 95% confidence interval excluding zero. Δ denotes the sex difference for that path, and Δ* denotes a significant sex difference (bootstrap 95% confidence interval for the difference excludes zero).

(C) Sex-difference estimates in the PC cohort (Δβ = female−male) for key mediation effects and path coefficients with bootstrap 95% confidence intervals. No Δβ effects were significant (all 95% CIs include zero), indicating no clear evidence for sex-dependent mediation effects in the PC cohort. This pattern should be interpreted cautiously given the smaller PC sample size, particularly in older age bins (**Supplementary Fig. S0**).

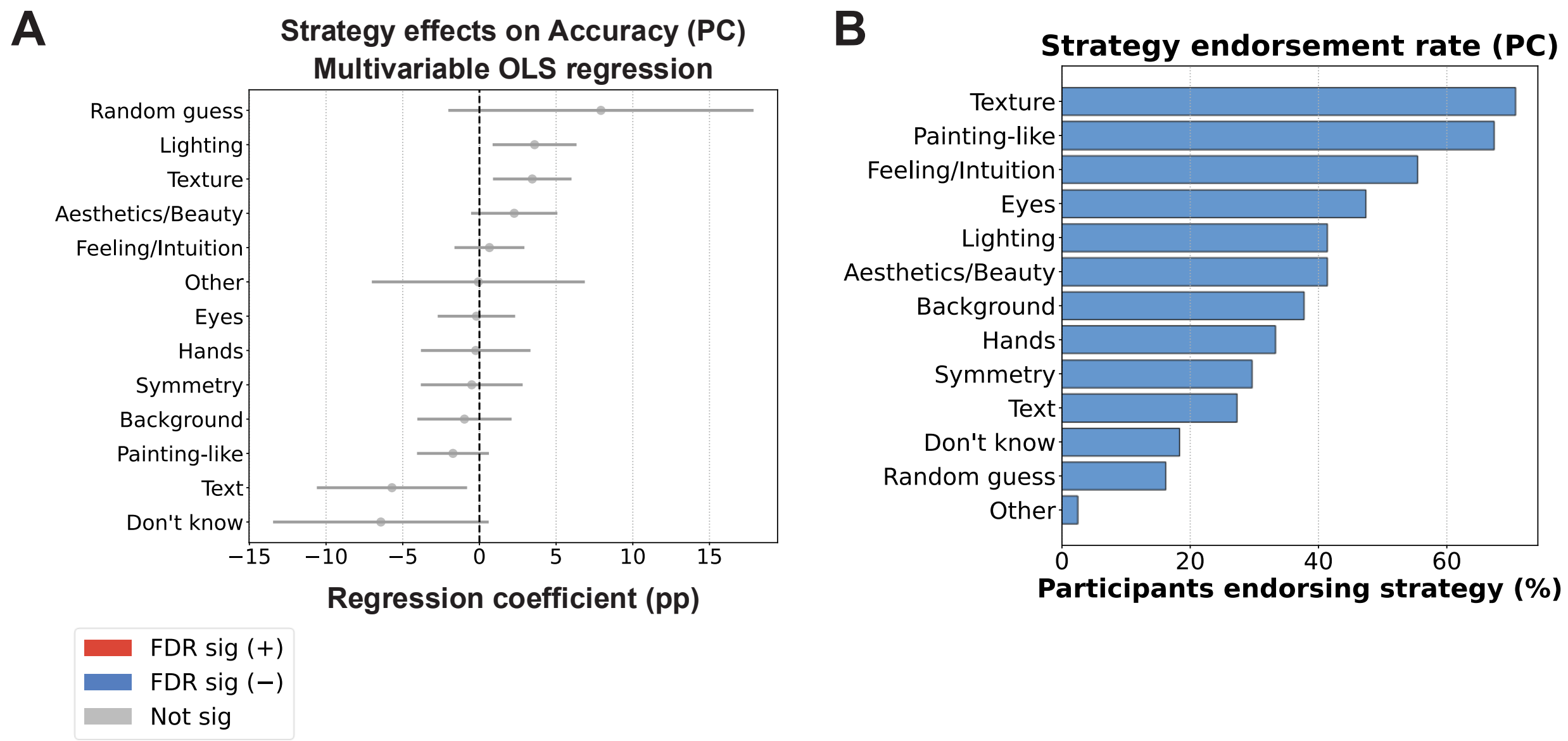

**Supplementary Figure S3. Strategy effects and strategy use in the PC cohort**

(A) Multivariable ordinary least squares (OLS) regression coefficients (percentage points) for strategy indicators predicting accuracy in the PC cohort, controlling for age, sex, and all other strategy indicators (95% confidence intervals computed using heteroskedasticity-consistent (HC3) robust standard errors; FDR correction across strategy terms). No strategy term survived FDR correction in the PC cohort.

(B) Strategy endorsement rate in the PC cohort (percentage of participants endorsing each strategy; multiple strategies could be selected per participant).

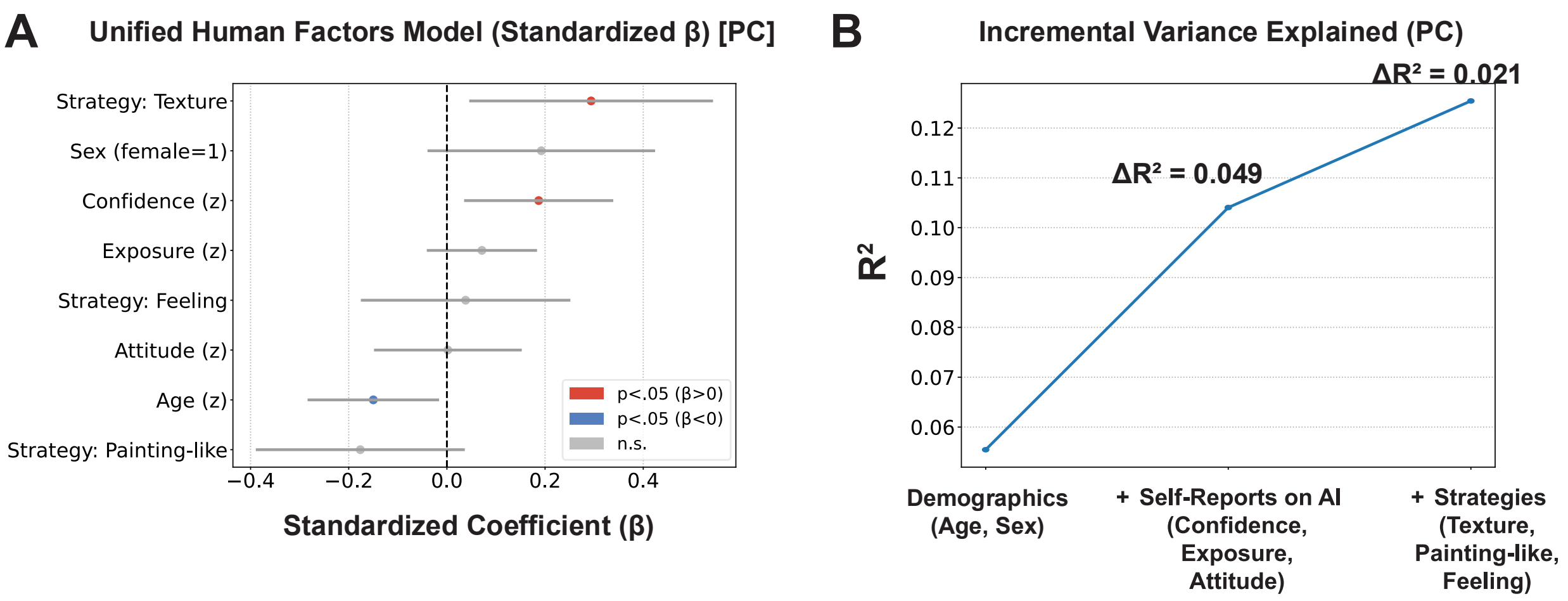

**Supplementary Figure S4. Unified human-factors model in the PC cohort**

(A) Standardized coefficients (β) from the unified regression model predicting discrimination accuracy in the PC cohort. Predictors were entered simultaneously and include demographics (age, sex), AI self-report measures (confidence, exposure, attitude), and a limited set of a priori selected strategy indicators (texture, painting-like appearance, feeling/intuition). Points show β and horizontal lines show 95% confidence intervals computed with heteroskedasticity-consistent (HC3) robust standard errors; colors indicate $p < .05$ (red: $\beta > 0$; blue: $\beta < 0$; gray: n.s.).

(B) Incremental variance explained ($R^2$) from nested models adding predictor blocks sequentially: demographics → AI self-reports → strategies. $\Delta R^2$ annotations indicate the added explanatory power of each block in the PC cohort. Note that the strategy terms in Panel A reflect a small a priori set used for the unified model, whereas the full strategy screen with FDR correction across all strategy terms is shown in **Supplementary Fig. S3**.

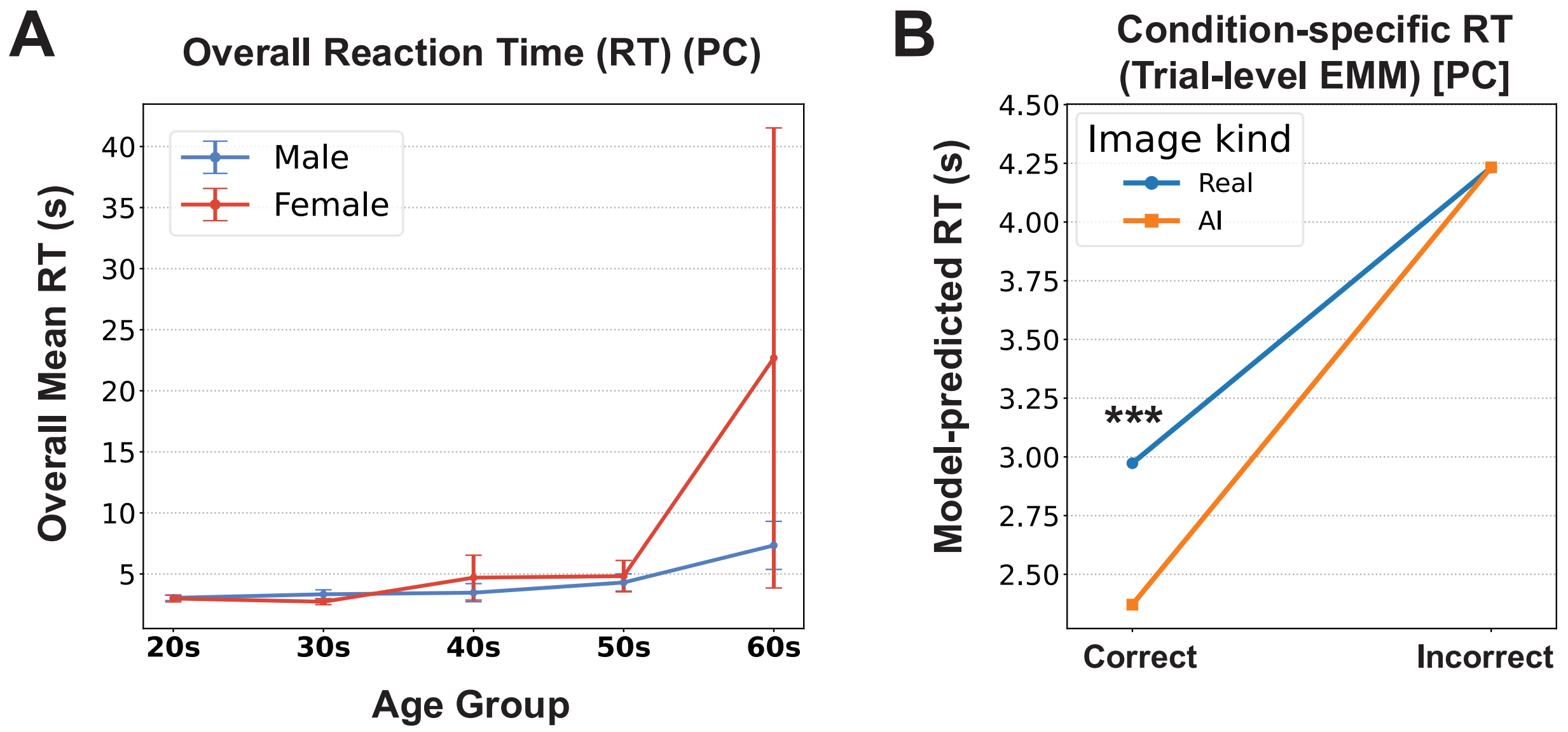

**Supplementary Figure S5. Reaction-time (RT) patterns in the PC cohort.**

(A) Overall mean RT by age group and sex in the PC cohort (mean ± SEM; RT in seconds).

(B) Condition-specific RT estimated from a trial-level linear mixed-effects model (LMM) of logRT with participant random intercepts, including fixed effects of Correctness × Image kind (Real vs AI), age, and sex; model-based estimated marginal means (EMMs) were back-transformed to seconds for visualization. Asterisks indicate a significant post hoc Real–AI contrast within the corresponding correctness condition (*** $p < .001$). Because the PC cohort contains few participants in older age bins—especially among females (**Supplementary Fig. S0**)—SEM estimates in panel A should be interpreted cautiously.

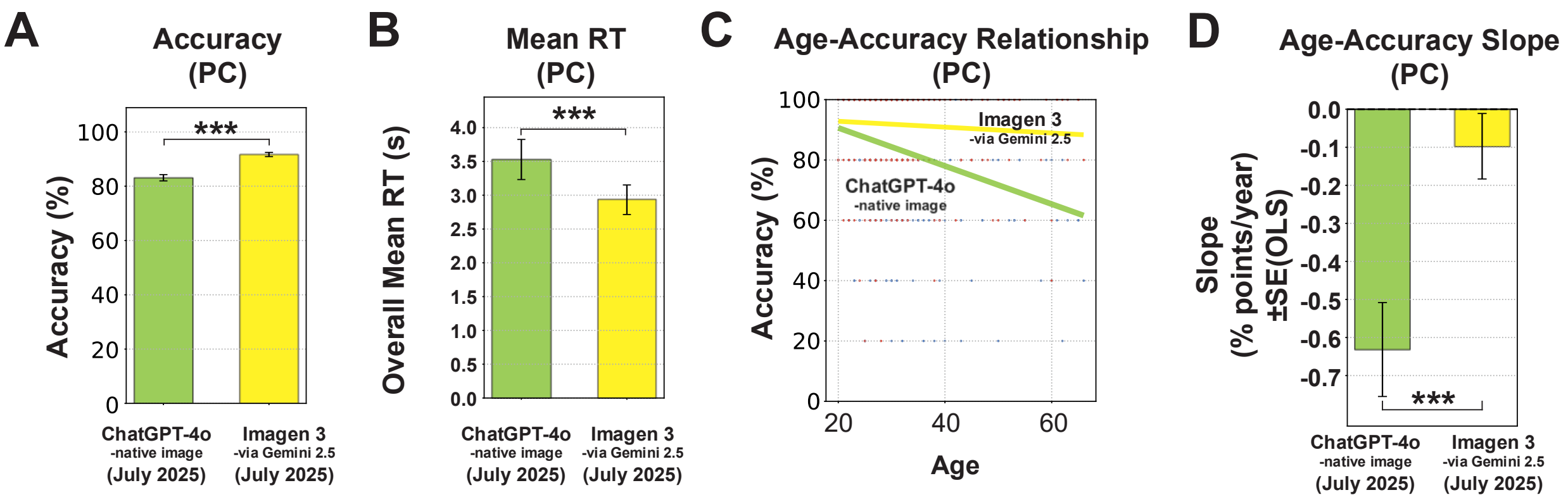

**Supplementary Figure S6. Generator-dependent differences in discrimination performance in the PC cohort (stimuli created July 2025).**

(A) Mean accuracy for AI portraits generated by ChatGPT-4o (native images) versus Imagen 3 (via Gemini 2.5) in the PC cohort (mean ± SEM; paired *t*-test, $p < .001$).

(B) Mean reaction time (RT, seconds; mean ± SEM) by generator (paired t-test, $p < .05$).

(C) Age–accuracy relationships plotted separately by generator with fitted linear trends.

(D) Estimated age–accuracy slopes (percentage points/year) by generator (ChatGPT-4o: −0.632 ± 0.123; Imagen 3: −0.097 ± 0.086; error bars show standard errors from ordinary least squares fits within each generator). Slope differences were tested using an age×generator interaction model with HC3 robust standard errors ($p < .01$).